Top marine predators track Lagrangian coherent structures

Emilie Tew Kai[1*], Vincent Rossi[2], Joel Sudre[2], Henri Weimerskirch[3], Cristobal Lopez[4], Emilio Hernandez-Garcia[4], Francis Marsac[1], and Veronique Garcon[2].

[1] IRD UR 109 THETIS, Centre de Recherche Halieutique, Avenue Jean Monnet - BP 171, 34203, Sète Cedex, France, * Corresponding author

[2]LEGOS, Centre National de la Recherche Scientifique, 31401 Toulouse Cedex 9, France

[3]Centre d'Etudes Biologiques de Chizé, Centre National de la Recherche Scientifique, 79360 Villiers en Bois, France

[4]IFISC, Instituto de Física Interdisciplinar y Sistemas Complejos, CSIC-Universitat de les Illes Balears, E-07122 Palma de Mallorca, Spain

- Email : emilie.tewkai@ird.fr

**Abstract**

Meso- and submesoscales (fronts, eddies, filaments) in surface ocean flow have a crucial influence on marine ecosystems. Their dynamics partly control the foraging behaviour and the displacement of marine top predators (tuna, birds, turtles, and cetaceans). In this work we focus on the role of submesoscale structures in the Mozambique Channel on the distribution of a marine predator, the Great Frigatebird. Using a newly developed dynamical concept, namely the Finite-Size Lyapunov Exponent (FSLE), we have identified Lagrangian coherent structures (LCSs) present in the surface flow in the Channel over a 2-month observation period (August and September 2003). By comparing seabirds' satellite positions with LCSs locations, we demonstrate that frigatebirds track precisely these structures in the Mozambique Channel, providing the first evidence that a top predator is able to track these FSLE ridges to locate food patches. After comparing bird positions during long and short trips, and different parts of these trips, we propose several hypotheses to understand how frigatebirds can follow these LCSs. The birds might use visual and/or olfactory cues and/or atmospheric current changes over the structures to move along these biological corridors. The birds being often associated to tuna schools around foraging areas, a thorough comprehension of their foraging behaviour and movement during the breeding season is crucial not only to seabirds' ecology but also to an appropriate ecosystemic approach of fisheries in the Channel.

\body

In the oligotrophic open ocean mesoscale and submesoscale oceanic turbulence, which spans spatiotemporal scales from one to hundreds of kilometers and from hours to weeks, strongly modulates the structure, biomass and rates of marine pelagic ecosystems. Eddies can stimulate the primary productivity (1, 2), affect plankton community composition (3-5) or play a significant role in exchange processes in the transitional area between the coast and offshore by transporting organic matter and marine organisms from the coast to the open ocean and vice versa (6). In view of the strong influence of eddies on physical and biogeochemical properties, it is not surprising that higher level predators concentrate around them, where prey can be found. In fact, all investigations on the relationship between eddies and top predators communities, using satellite imagery observations, have evidenced strong ties between them (7, 8). Upper predators particularly used the boundary between two eddies (9 -12). The key point is that interactions between eddies generate strong dynamical interfaces (13) and make them a complex and energetic physical environment. In these interfaces the energy of the physical system is available to biological processes, increasing the trophic energy of the biological system (8). Eddies and associated structures have therefore a crucial ecological significance especially in tropical and sub-tropical regions, characterized by low mixing during winter inferring weak supply of nutrients to the photic zone (11).

Most previous works dealing with the influence of eddies on top-predator distribution show the necessity to concentrate on submesoscale (below 10 km) to fully appreciate the role of eddy-eddy interfaces on biological production (11). Many different studies confirm that submesoscale tracer patches and filaments are strongly related to interactions between mesoscale surface eddies (1, 14). Despite this, studies on top predators using remote sensing have only used Sea Surface Height (SSH) as an indicator of eddy activity, which does not resolve sub-mesoscale structures such as filaments, where production should be concentrated. In addition, a fundamental question remains: how top predators can find these zones of higher productivity? This is particularly difficult to understand for central place foragers such as seabirds that breed on land but have to do continuous return trips between feeding zones and the colony where they care for their chick or egg. The additional difficulty in the case of eddies is that the location of production zones moves continuously.

In the West Indian Ocean, the Mozambique Channel (hereafter MC) can be considered as a natural laboratory to study the interactions between biological and physical processes at mesoscale in oligotrophic areas (sub-tropical region) due to the transient activity of eddies. Indeed mesoscale dynamics of the Mozambique Channel has been well described by previous works using remote sensing data, modelling and *in situ* observations (15-17). Mesoscale activity is dominant in two areas, the central part of the MC and south of Madagascar (17, 18). Weimerskirch et al. (10) have shown the main role of mesoscale eddies on the foraging strategy of the Great Frigatebirds. These birds fly hundreds or thousands of kilometres from the colony in a few days and spend their entire foraging trips in flight, being unable to sit on the water or enter the water column. Bird's pathways are preferentially associated with eddies in the MC during their long trips and especially with the edge of eddies, avoiding their core (10). However it is not clear where they exactly forage in the eddy system and whether and how they locate the zones of high production. The aim of our study here is first to describe the fine scale activity occurring at the edge of eddies and other submesoscale structures, and quantify the role of these on a top predator's foraging movements. Finally, we will try to understand how and why these predators might locate these structures.

For the physical environment, we have used horizontal velocity fields computed from satellite altimetry products (19). We have applied to them a recently developed Lagrangian technique, the Finite Size Lyapunov Exponent (FSLE), which allows computing, from marine surface velocity field data, mixing activity and coherent structures that control transport at specified scales (20). FSLEs measure how fast fluid particles separate to a specified distance. Lagrangian coherent structures (LCSs), e.g. transport barriers, filamental structures or vortex boundaries, are identified as ridges (locations containing the maximum values) of Lyapunov exponent fields (21-24). Dispersion rates of tracer particles can be calculated by integrating trajectories towards the future (forward direction) or towards the past (backward), giving rise to two different quantifiers, $FSLE_f$ and $FSLE_b$, respectively, containing complementary information (see Methods section). Ridges of $FSLE_b$ attract neighboring trajectories whereas $FSLE_f$ repel them. This is why we call them *attracting and repelling LCSs*, respectively. Sometimes, especially for plotting, it is convenient to write $FSLE_b$ and $FSLE_f$ as having negative and positive values, respectively, and expressions such as $|FSLE|$ refer simultaneously to both types of exponents. For the marine top predators, we have used Argos positions of Great Frigatebirds from the colony in Europa Island in the MC during August-September 2003. Additional details are given in the Material and Methods section.

In this paper, we test if seabirds' positions during their foraging trips are related to dynamical structures. This is performed in different contexts: during short and long trips, day and night, and during the outward part of their foraging trip and return part back to the colony. We finally discuss which foraging strategy these top predators might use to locate prey patches.

## RESULTS

### Seabirds' locations during trips and FSLE fields

We compare here the locations of the LCSs identified as ridges in FSLE maps, and measured bird positions during August-September 2003. We will see that the latter are not random but correlated with the former.

First, Figure 1 shows Argos positions of Great Frigatebirds during long trips (black points) and short trips (red points), between August 18 and September 30, 2003. Locations of seabirds during long trips superimposed on FSLEs fields (September 24 to October 6, 2003), are shown in Figure 2. During the week of September 24, bird 11377 (green circles) is located on high $FSLE_b$ values (the attracting LCSs), as well as location of bird 16255 (blue circles). Positions of bird 8023 (red circles) seem to be linked with fluid repelling structures (the ridges of $FSLE_f$) instead. For bird 8023, at the beginning of the travel, the trajectory is rectilinear in the north-east direction and then follows the repelling mushroom-like structures. Foraging patches (triangles), where birds reduce flying speed, seem to exhibit the same distribution than the birds' moving positions. During the week of October 6, movements of bird 8023 are mostly on repelling structures (Fig.2, d) as during the week of September 24, and perhaps also on some attracting structures. The important point is that any of both types of LCSs is more visited than locations outside. Positions of bird 19827 (magenta circle) are well superimposed on fluid attracting structures (ridges of $FSLE_b$) but not on repelling ones. These two examples of the overlay of seabirds' moving and foraging positions on FSLE fields during long trips show that the locations of birds tend to overlay on LCSs either on attracting (Fig.2, a-c ) or repelling ones (Fig.2, b-d).

To put the above observations in quantitative form, we defined a threshold defining a significant presence of LCSs: $|\mathrm{FSLE}| > 0.1\ d^{-1}$ . It corresponds to mixing times smaller than one month. This value is chosen since it is a typical value for Lyapunov exponents in different areas of the globe (14, 20) and because regions where the Lyapunov exponents are larger already have the shape of one-dimensional lines (see Fig 2). The distributions of FSLEs in the whole MC and central part, and in areas crossed by seabirds were tested for conformity to the normal distribution using the Kolmogorov–Smirnov sample test and they all are clearly non-normal. Histograms of relative frequency of FSLE in the whole MC, central part and on areas visited by seabirds are shown in Figure 3. In the whole MC and central part, Lagrangian structures detected by $|\mathrm{FSLE}|>0.1$ $\mathrm{day}^{-1}$ represent a minority of locations, occupying 30% or less of the total area. However in areas crossed by frigatebirds more than 60% of the birds are on LCSs. Five Kolmogorov-Smirnov 2-sample tests (KS2) comparing the distributions of FSLEs in the whole MC and in the central part with the distribution of FSLEs on areas visited by seabirds during long and short trips were performed. The tests confirmed that distributions of FSLEs in areas crossed by seabirds are highly different from those found over the whole area and central part ($p<0.0001$ for both long and short trips). Distribution patterns provide clear evidence that Great Frigatebirds are not randomly distributed throughout the FSLE range (both backward and forward) and that seabirds move over specific areas rich in LCSs, despite the area occupied by LCSs is small. Close to 2/3 of the birds positions are on LCSs, despite that only 30% or less of the whole area or of the central part (see Fig. 3) contain high |FSLE| and are then occupied by LCSs. These numbers are further checked by chi-square analyses using the one tailed G-test for Goodness of Fit (Log-Likelihood ratio) which show clearly that there are significant differences between positions of birds on LCSs and on other structures (Table 1) (G-test, $p<0.001$): this

confirms again that seabirds' positions are located more on LCSs (|FSLE|>0.1 day $^{-1}$) than outside during long and short trips, despite the small area occupied by LCS (Fig. 3). An additional test checking the relation between birds' positions at a given week t and the LCSs computed for that week and for the following ones, t+1, t+2, …, t+9, is described in SI. The association of birds' tracks and LCSs, measured by the significance of a G-test, is highest for the LCSs of the week t and decreases with the time lag to the other weeks ($p_{t+1}$=0.81 > $p_{t+3}$=0.19 > $p_{t+5}$=0.12) (Supporting Information [SI], Table S1).

**FSLE distributions over different types of flights**

We performed several statistical tests to see if there are statistically significant differences among travel/foraging locations, outgoing/return trips, and day/night flights. Boxplots of FSLEs on seabirds' positions during long and short trips are presented in Figure 4. The range of variation of FSLE is clearly more dispersed during long trips than short trips and the median between both kinds of trips is similar. Furthermore, distributions are clearly different between long and short trips as confirmed by a KS-2 samples test (p<<0.001). Indeed, 65.9 % of seabirds' positions during long trips and 56 % during short trips are on LCSs (Table 1). During long trips, Great Frigatebirds forage during a longer time, and so cover a larger range of variation of FSLE values than during short trips. One tailed G-test for Goodness of Fit confirms that there is a difference between the number of seabirds' locations on FSLE ridges and outside the ridges (Table 1) (G=30.613; p=0.001; df (degrees of freedom)=10 for long trips and G=32.057; p<<0.001; df=6 for short trips).

KS-2 tests show that the distribution of the birds between attracting and repelling LCSs display no statistically significant difference during long trips (p>0.05) but differ during short trips (p<0.01). During short trips birds follow more the attracting LCSs than the repelling ones. The analyses clearly demonstrate that seabirds follow the FSLE ridges during their foraging trips, but mostly during long trips than during short trips. This result underlines the probable difference between the Great Frigatebirds behaviour during long and short trips.

Boxplots of FSLE show that patterns of distribution of FSLE are not very different between flying and foraging positions (SI, Fig. S1). Distributions of FSLEs are statistically similar for foraging and crossed areas (KS-2 test, $p=0.29$ for long trips and $p=0.51$ for short trips), but differ from FSLE distribution in the whole area (KS test $p<0.0001$). During long trips 69.6% (resp. 61.8% during short trips) of seabirds' positions during flying and 62% (resp. 66.7% during short trips) during foraging are on LCSs (SI Fig.1). During flying and foraging seabirds split almost equally between repelling and attracting structures (G-test $p>0.05$) (see SI, Table S2). All of this indicates that seabirds seem to prefer being on ridges of FSLE both for travel and foraging.

We have also investigated for differences in seabirds' distributions in relation to FSLEs between the outward and return part of the trip (see SI, Fig. S2a, c). KS-2 test shows that there is no significant difference of seabirds' distribution during long trips (KS-2 $p>0.01$) and during short trips ($p>0.05$), between the outward and return parts of the trip. For all types of trips (short and long), there is no significant difference of seabirds' positions, either on repelling or attracting flow structures, during the outward and return parts of the trip (G-tests $p > 0.05$) (see SI, Table S3).

Great Frigatebirds feed mainly during daytime (10). We therefore examined whether we could identify differences between day-time and night-time distribution of seabirds. Boxplots of seabirds' distribution on FSLE between day and night show that patterns of distribution of FSLEs are similar during day and night during short (SI, Fig. S2b) and long trips (SI, Fig. S2d). The range of variation of FSLE during long trips is however more dispersed at night than during short trips. KS-2 test shows that there is no significant difference between FSLE distributions visited by birds during day and night ($p>0.05$ during long or short trips). The probability for the frigatebirds to fly over attracting or repelling structures during day and night is statistically similar (G-tests $p>0.05$) for long trips but may be different for short trips (G-test $p=0.025$) (SI, Table S3). During daytime short trips, seabirds may follow more the attracting structures than the repelling ones.

## DISCUSSION

As eddies affect all stages of the marine ecosystem, they are determinant for the triad "enhancement-concentration-retention" identified by Bakun (25, 8). From upwelling-driven processes at the centre of cyclonic eddies (1, 2), or from other processes at the boundaries between eddies (13), local enrichment and new production have been observed. The cyclic circulation in vortices produces also retention of larvae and other planktonic organisms in their core, whereas concentration occurs in the convergence zones located at the boundary between them, which are detected by FSLEs.

Transport barriers and filament generation by interaction between eddies induce horizontal and vertical biogeochemical and biological enhancement (13). Finite Size Lyapunov Exponents seem very well-suited to detect such transport barriers, vortex boundaries, and filaments at meso- and submesoscale (20, 26) and to study the link with the ecological behaviour of marine top predators. However, a word of caution is required about the spatial resolution we used. Indeed, the FSLEs are computed from satellite altimetry products (19) with a spatial resolution of 1/4 of a degree interpolated here onto a 1/40 of a degree grid. This interpolation might induce some bias in the data. However FSLEs, because of the averaging effect produced by computing them by integrating over trajectories which extend in time and space, are rather robust against noise and uncertainties in velocity data (26, 27) (see also SI). The velocity field used here has been validated and the correlation with velocities from Lagrangian drifting buoy data in the MC was satisfactory (see SI). Furthermore, Argos positioning of birds is not of equivalent quality. Some positions have a margin of error of a few hundred metres, while others have an error margin of more than one kilometre. Definite improvements would be to reduce interpolation by using an original higher resolution velocity field and to obtain more precise birds' locations.

In the central part of the Mozambique Channel, it is known that the boundary of eddies is very energetic and allows the aggregation of top predators foraging, especially Great Frigatebirds (10), which preferentially stay in this part of the channel. So far it was believed that Great Frigatebirds used edges of eddies mainly for food because these areas are rich in forage species and associated top predators (especially tuna and dolphins, (28)). Superimposing Great Frigatebirds's positions on FSLE fields shows that their spatial distribution is linked to eddies, and more generally to the different types of LCSs. And not only for foraging but also for travelling. Observations are in agreement with the histograms and Kolmogorov-Smirnov tests, which demonstrate that seabirds are not randomly distributed in relation to attracting and repelling LCSs.

However, analysis of location of seabirds during long and short trips shows that the percentage of positions on LCSs is different between both kinds of trips (Table 1). During long trips, birds seem to take full measure of the LCSs while on short trips they do not take full advantage of them. This difference between long and short trips is probably due to the behaviour of seabirds. During short trips, birds have to bring food frequently to their chick so they feed in areas where preys are easily accessible, close to Europa Island. They used preferentially attracting structures during daytime, probably because these structures are conductive to the aggregation of preys. During long trips, birds avoid areas near Europa Island probably because the foraging yield is less rich than that of more distant waters, and/or because of strong interspecific competition near the island (10). However, birds preferentially follow the LCSs in both cases.

In addition, seabirds follow LCSs not only for their foraging but also for their travelling movements. The distributions of FSLEs during the outward and inbound journeys to the colony indicate that they exhibit the same flying behaviour before and after their foraging activity. Furthermore, the fact that the distribution of visited FSLEs is identical during day and night indicate that they are able to use these LCSs to move during periods of darkness. Frigatebirds move continuously during day and night at an average altitude of 200 m, and never completely stop moving when they forage, but they come to the sea surface to eat only during day-time (10). If they used these structures only for food availability, then the distribution of FSLEs for areas crossed by birds should be different between day and night. This is not the case. This means that frigatebirds do not go to FSLEs ridges only to forage but that they follow them most of the time as cues to eventually find prey patches there.

It is relatively easy to understand why the attracting LCSs could be places for prey accumulation, since horizontal flow will make passively advected organisms close to these lines to approach them. More puzzling is to understand the role of the repelling LCSs, which are also preferred locations for the frigatebirds. First we should mention that at the vortex edges, lines of the attracting and the repelling types are very close and nearly tangent. Thus, it may be the case that birds' positions located at repelling lines are simultaneously located also on attracting ones: in SI we explain that a position is said to be on a LCS if it is closer to it than 0.025 degrees. Thus, if the attracting and repelling LCSs are close enough, the same bird position may be attributed to both structures. We have checked that, among the 30.2% of bird positions which were found on repelling coherent structures, 53.7% of them were in fact visiting both structures, and thus the interpretation is that they are associated to vortex edges (or to other structures in which both types of lines are tangent). For the remaining fraction which does not seem to be associated to these edges, we believe that the three-dimensional dynamics of the flow close to these structures gives the clue for their association to birds' positions. Note that FSLE values have been calculated on the basis of the two-dimensional surface flow, and the FSLE methodology identifies these regions as places of filament and submesoscale structure formation by horizontal advection. But there is growing evidence (29,30) of strong links between submesoscale structures from different origins and vertical motions. Thus, in an indirect manner, the calculated LCSs may be indicating the places in the ocean where vertical upwelling and/or downwelling of nutrients and organisms could occur. This is obviously important for the birds, and may explain why they prefer to fly and to forage on top of them. The role of these LCSs on the biological activity is rather complex and may vary depending on the area and scale of study. For instance, (31) found an inverse relationship between mixing activity (high FSLEs) and phytoplankton stocks in very productive areas such as coastal eastern boundary upwelling.

The above arguments linking LCSs and vertical motion can be more easily justified for the attracting LCS case, because the vorticity involved in the interaction between vertical and horizontal motion will tend also to be aligned with these structures (30). But we note that in flows consisting on slowly moving eddies, we are close to the so-called integrable situation in which a large proportion of tangencies between attracting and repelling structures is expected (as indeed observed). As a consequence, it may happen that a bird starts a trip by following an attracting LCS, loses its surface signal, and finds itself on top of a repelling one simply by continuing its previous path in a more or less straight way.

We stress, however, that all explanations we give to the observed relationship between LCSs and bird paths contain a number of hypothesis which need additional research.

Besides, one may ask how can frigatebirds “follow” the LCSs during day and night. Several hypotheses can be put forward:

- First, because frigatebirds use atmospheric currents, especially to gain altitudes by soaring, and then glide over long distances (32), we can suppose that the coupling between the ocean and the atmosphere at meso and submesoscale generates atmospheric currents followed by seabirds. Indeed some authors (33-36) underline the role of local air–sea feedbacks arising from ocean mesoscale features. For example Chelton et al. (36) showed that an ocean-atmosphere coupling is observed in the California Current System during summer. They conclude that SST fronts generated by mesoscale activity (eddies and upwelling) have a clear influence on the perturbation of summertime wind stress curl and divergence. In the Mozambique Channel, mesoscale eddies and their interaction would force the atmosphere and generate air-current favourable to Great Frigatebirds that might take advantage of the wind to fly in spending the least possible energy.

- Second, we cannot exclude that birds may follow visual or, more likely, olfactory cues. Foraging behaviour of seabirds is complex and results from a number of behavioural parameters such as sight, smell (37, 38), memory effect (39) and environmental parameters: chlorophyll concentration (10), or wind speed and direction. Nevitt et al. (40) suggest that seabirds use olfaction to track high concentrations of odour compounds such as dimethyl sulphide (DMS) and sight when they locate prey patches. The use of models of odours transport suggests that olfaction plays a role in foraging behaviour (40). Structures detected using FSLEs are dynamical and, as mentioned above may induce vertical mixing favourable to phytoplankton enhancement (41, 42) and their patchy distribution. The grazing of phytoplankton by zooplankton induces the production of DMS (43) which is very attractive for different species of seabirds (44). Even if there is no study on the role of olfaction on Great Frigatebirds foraging behaviour, we can hypothesize that they use olfaction to detect DMS and productive areas and find food patches. The interaction between the ocean and the atmosphere at sub-mesoscale and wind may allow the dispersion of the DMS or other odours and favour their detection by seabirds that follow LCSs until they see a patch prey. These LCSs could be viewed as moving habitat facilitating movement of seabirds. Indeed frigatebirds might use these odourful corridors to move between food patches with efficacy.

Whatever is the cue used by frigatebirds to locate and follow these Lagrangian coherent structures, our results provide the first evidence that a top predator tracks these FSLE ridges to locate food patches. It allows us to better understand how top predators search preys, and why they are able to concentrate precisely at LCSs. Since these structures are mobile, a simple memory is not sufficient for a central place forager to return to a productive prey area. Predators could thus take a general bearing where eddies are likely to be found (e.g. to the northwest in the MC for a colony located in the central MC) and then move until they cross a FSLE ridge, that they will follow until they encounter a prey patch. Because they are unable to sit on the water, frigates are often in association with sub-surface top predators to forage. We can suppose that if frigatebirds track LCSs to locate preys, it is possible that they are associated to tuna schools around foraging areas (10). Thus understanding the rationale behind their localization is crucial in seabird's ecology but also in the detection of the presence of tuna schools. This kind of multidisciplinary approach opens up interesting prospects in the management of ecosystems and fisheries and can be useful in the ecosystemic approach to fisheries, especially to better characterize temporary tuna habitats in the Mozambique Channel.

Future work is to identify the responsible mechanism by which an aerial predator may spot and follow LCSs.

## MATERIAL AND METHODS

In this part we provide a brief overview of the methodology; further details for each section are explained in the Methods in *SI Text*.

### Great Frigatebirds

Europa (22.3° S, 40.3° E) is one of the two colonies (with Aldabra) of Great Frigatebirds in the West Indian Ocean. The island is located in the central part of the Mozambique Channel. Great Frigatebirds have the ability to undertake long range movements out of the breeding season (10) but they behave as central place foragers when breeding. Their diet is composed essentially of flying-fish and Ommastrephid squids (10), but Great Frigatebirds are also kleptoparasits meaning they can steal preys from others. One of their particularities is that they cannot wet their feathers nor dive into the water to feed. They forage mainly through association with tuna and dolphins schools, which bring prey to the surface.

To track movements of frigatebirds, 8 birds were tracked with satellite transmitters and altimeters between August 18 and September 30, 2003, resulting in 1864 Argos positions. The mean time between each position is 0.07 days, with a minimum of 0.001 days and a maximum of 1.1 days. All seabirds positions from a given week were collocated on the time and space grid were the FSLEs were calculated (with 0.025° resolution).

### Lagrangian coherent structures by Finite Size Lyapunov Exponents

*FSLE method*

Oceanic variability in surface velocities is not probably sensed directly by Great Frigatebirds, but indirectly via transported substances. This calls for a Lagrangian perspective of the problem. Thus, we quantify horizontal transport processes and Lagrangian coherent structures by the Lagrangian technique of the Finite Size Lyapunov Exponents (FSLE) (45), which is specially suited to study the stretching and contraction properties of transport in geophysical data (20). Due to its Lagrangian character, FSLEs describe submesoscale details which cannot be detected by other means, like the inspection of the Sea Level Anomaly maps of the marine surface.

The calculation of the FSLE goes through computing the time, $\tau$, at which two tracer particles initially separated at a distance $\delta_0$, reach a final separation distance $\delta_f$, following their trajectories in the marine surface velocity field. At position x and time t the FSLE is given by:

$$\lambda\left(x,t,\delta_0,\delta_f\right)=\frac{1}{\tau}\log\left(\frac{\delta_f}{\delta_0}\right) . \quad (1)$$

We follow the trajectories for 200 days, so that if $\tau$ is larger than this, we define $\lambda=0$. It is clear that the FSLEs depend critically on the choice of two length scales: the initial separation $\delta_0$ and the final one, $\delta_f$. $\delta_0$ has to be close to the intergrid spacing among the points x on which the FSLEs will be computed (20). In our case we calculate FSLE on all the points of a latitude-longitude grid with a spacing of $\delta_0$ = 1/40°=0.025°. On the other hand, since we are interested in mesoscale structures, $\delta_f$ is chosen as $\delta_f$ = 1°, i.e., separation of about 110 km. In this respect, the FSLE represents the inverse time scale for mixing up fluid parcels between the grid and the characteristic scales of the Mozambique Channel eddies. Maps of FSLE are calculated weekly. An alternative to FSLE are the finite-time Lyapunov exponents (FTLE) (22, 46). At the scales and parameters we are working no significant differences are expected for the locations of LCS by any of the two methods.

The time integration of the particle trajectories can be performed in two different ways: forward and backward in time. For the backward in time computation, maximum values of FSLE organize in lines which are good approximations to the so called *unstable manifolds of hyperbolic points*, which for our purposes are lines towards which neighboring fluid trajectories, while escaping from hyperbolic points, approach at long times (20, 23, 24). In consequence they are called attracting LCSs. FSLEs computed integrating trajectories towards the future, i.e. forward-in-time, take large values on lines (stable manifolds) from which neighbouring trajectories appear to be repelled (repelling LCSs). These lines of maximum separation or convergence rates, or “ridges”, delineate fluid domains with quite distinct origin and characteristics. Such lines strongly modulate the fluid motion since when reaching maximum values, and they act as transport barriers for particle trajectories thus constituting a powerful tool for predicting fronts generated by passive advection, eddy boundaries, material filaments, etc. In different sets of papers (20, 26, 27, 31, 42), it has been demonstrated the adequacy of the FSLE to characterize horizontal mixing and transport structures in the marine surface, as well as its usefulness when correlating with tracer fields like temperature or chlorophyll.

## ACKNOWLEDGEMENTS

Ph.D. fellowship for E.T.K has been provided by the Institut de Recherche pour le Développement and the University Pierre and Marie Curie. PhD financial support for V.R has been provided by the Direction Générale de l’Armement. The LEGOS Contribution is supported through CNES funding. IFISC contribution is supported by MICINN and FEDER trough project FISICOS (FIS2007-60327), and by CSIC through PIF project OCEANTECH. HW's contribution was supported by the REMIGE project funded by Agence Nationale de la Recherche (ANR 2005 Biodiv-011). We acknowledge the two anonymous reviewers for their helpful comments on the manuscript.

**Legend of figures**

Figure 1: Argos locations of Great Frigatebirds during long trips (black points) and short trips (red points) in the Mozambique Channel, between August 18 and September 30, 2003. The green point denotes Europa Island.

Figure 2: Overlays of seabirds' position on FSLE maps. Left panels (A and C): Backward integration in time for FLSE computation ($d^{-1}$). Right panels (B and D): forward integration in time ($d^{-1}$). A and B, week of September, 24, 2003. C and D, week of October, 6, 2003. Circles represent seabirds trajectory and triangles foraging patches. Each color of points represents the tag of a different bird (red, tag 8023; blue, tag 16255; green, tag 11377; magenta, tag 19827).

Figure 3: Histograms of relative frequency of FSLEs with percent of attracting (ALCSs) and repelling LCSs (RLCSs). Positive values refer to $FSLE_f$ and negative to $FSLE_b$. A) areas crossed by seabirds (long and short trips); B) in the whole MC and C) in the central part (16°S-24°S/30-45°E)

Figure 4: Box plots of the distribution of FSLEs during short and long trips. The upper and lower ends of the center box indicate the 75th and 25th percentiles of the data; the center of the box indicates the median. Suspected outliers appear in a box plot as individual points + outside the box. Dotted lines represent the threshold for detection of LCSs.

## Caption of table

Table 1: Absolute frequency of seabirds' positions on LCSs and on no Lagrangian structures for long and short trips per week and result of the G-test for Goodness of Fit. *Alpha 5%.*

| | All trips | | Long trips | | Short trips | |
|---|---|---|---|---|---|---|
| week | **LCSs: \|FSLE\|>0.1 $day^{-1}$** | **\|FSLE\|<0.1 $day^{-1}$** | **LCSs: \|FSLE\|>0.1 $day^{-1}$** | **\|FSLE\|<0.1 $day^{-1}$** | **LCSs: \|FSLE\|>0.1 $day^{-1}$** | **\|FSLE\|<0.1 $day^{-1}$** |
| 1 | 38 | 9 | 19 | 7 | 19 | 2 |
| 2 | 78 | 40 | 55 | 12 | 23 | 28 |
| 4 | 208 | 85 | 147 | 54 | 61 | 31 |
| 5 | 167 | 109 | 137 | 84 | 30 | 25 |
| 6 | 120 | 77 | 89 | 51 | 31 | 26 |
| 7 | 79 | 55 | 72 | 32 | 7 | 23 |
| 8 | 53 | 34 | 53 | 34 | - | - |
| 9 | 61 | 59 | 61 | 59 | - | - |
| 10 | 55 | 31 | 45 | 24 | 10 | 7 |
| 14 | 35 | 12 | 35 | 12 | - | - |
| 15 | 10 | 5 | 10 | 5 | - | - |
| % | 63.7 | 36.3 | 65.9 | 34.1 | 56.0 | 44.0 |
| | *G Test (Log-Likelihood ratio)* | | *G Test (Log-Likelihood ratio)* | | *G Test (Log-Likelihood ratio)* | |
| N | 1420 | | 1097 | | 323 | |
| k | 11 | | 11 | | 7 | |
| df | 10 | | 10 | | 6 | |
| G | 28.119 | | 30.613 | | 32.057 | |
| p | 0.00173 | | 0.001 | | 0.000 | |

*One tailed tests.Ho: Seabird positions share equally LCSs (|FSLE|>0.1 $day^{-1}$ and on no LCSs.*

TABLE 1

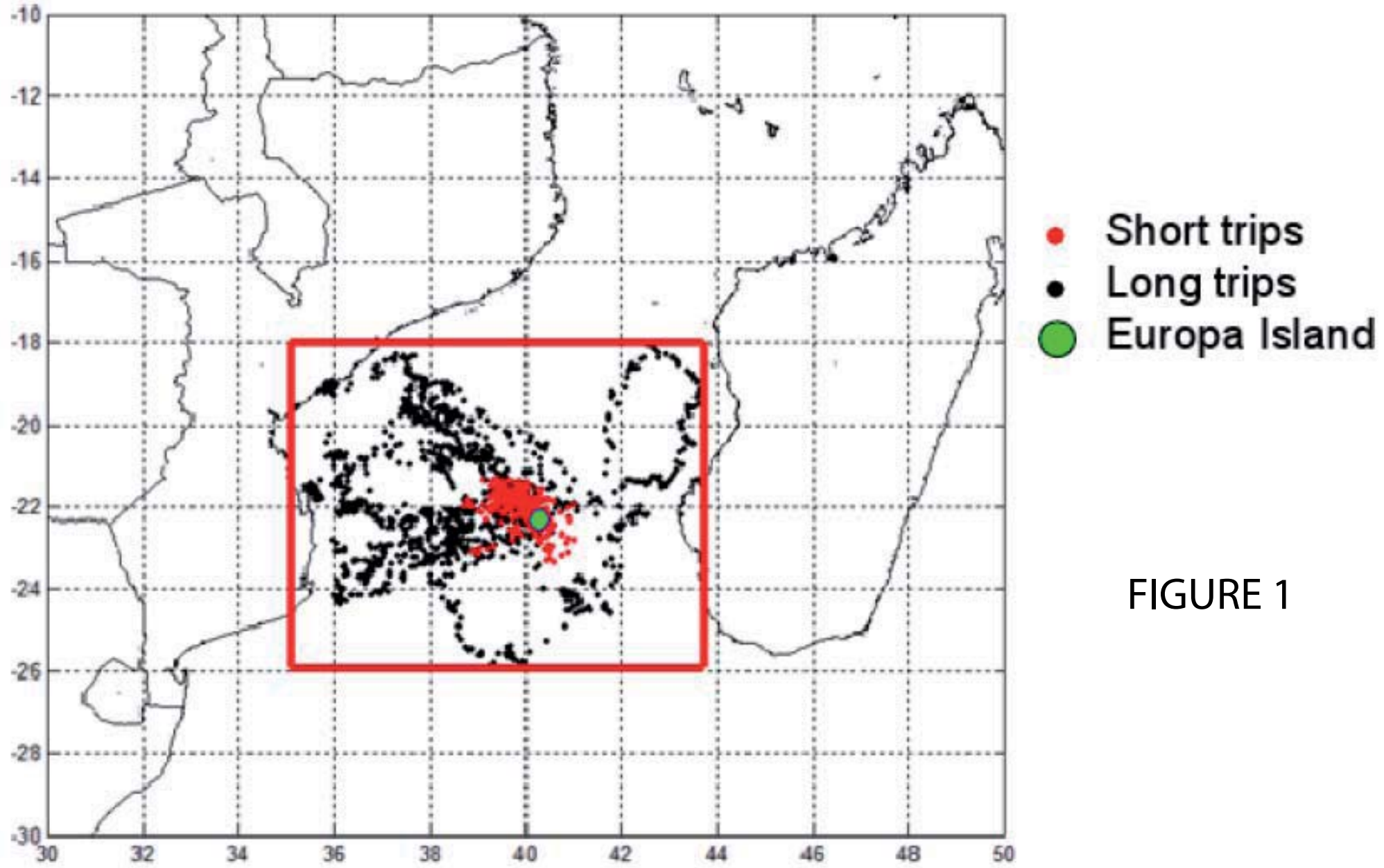

FIGURE 1

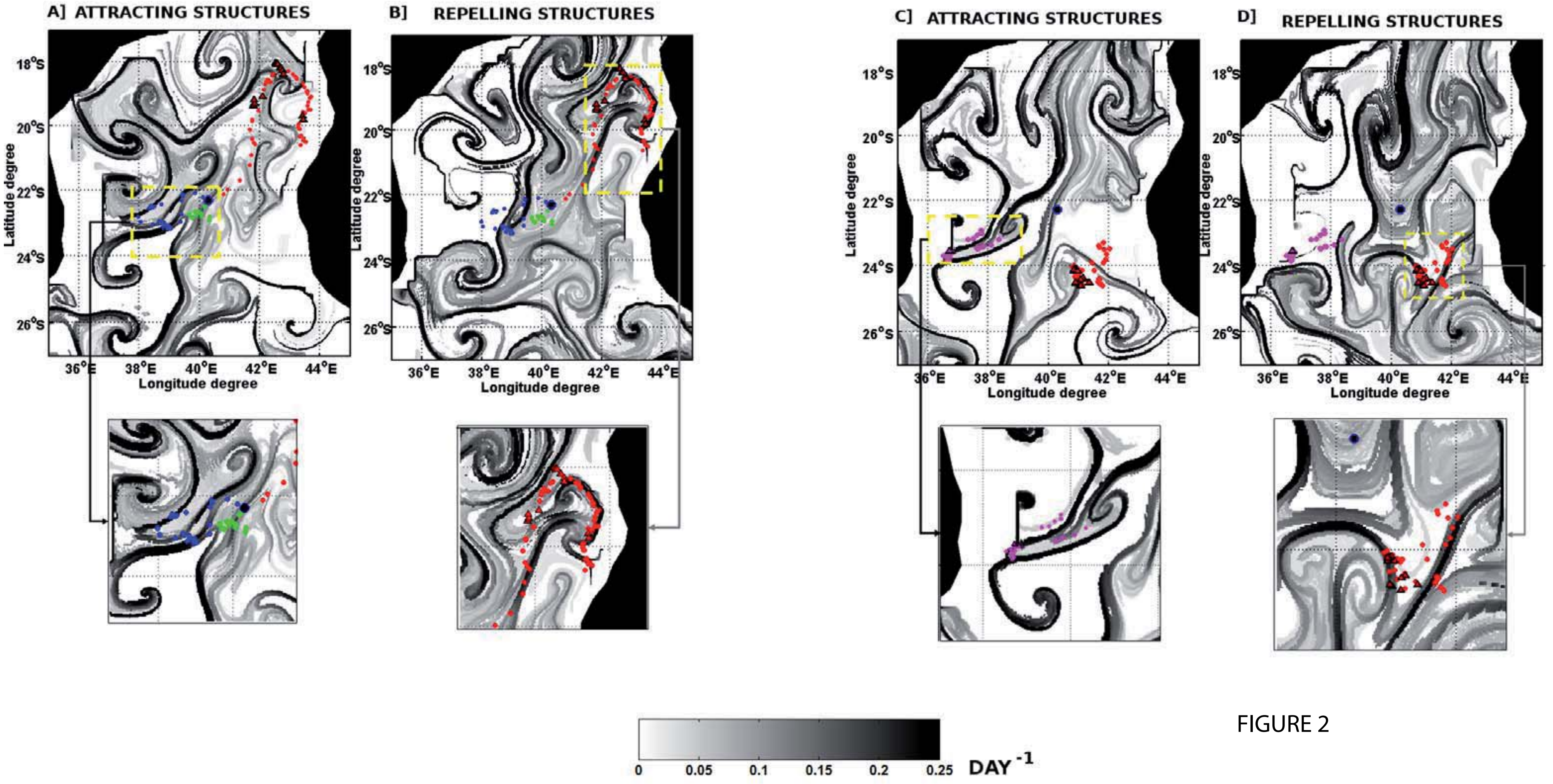

FIGURE 2

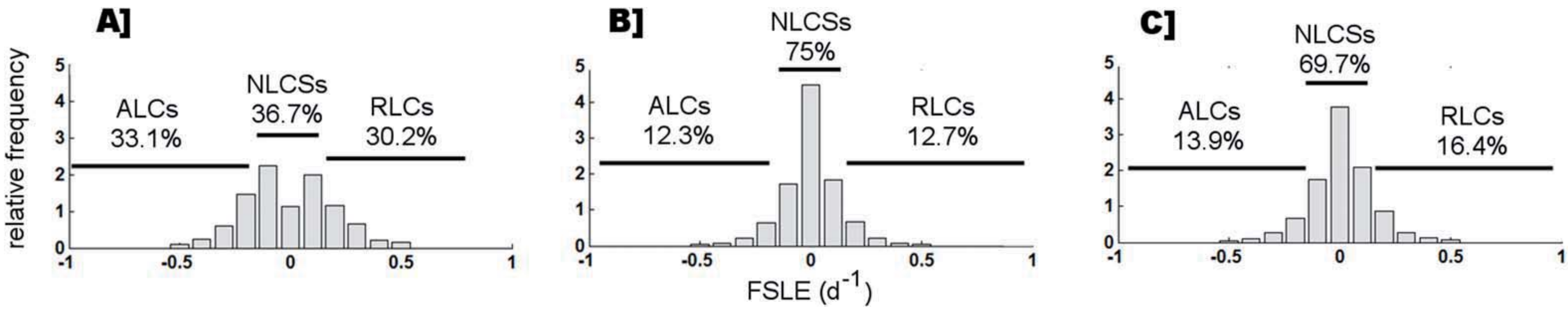

FIGURE 3

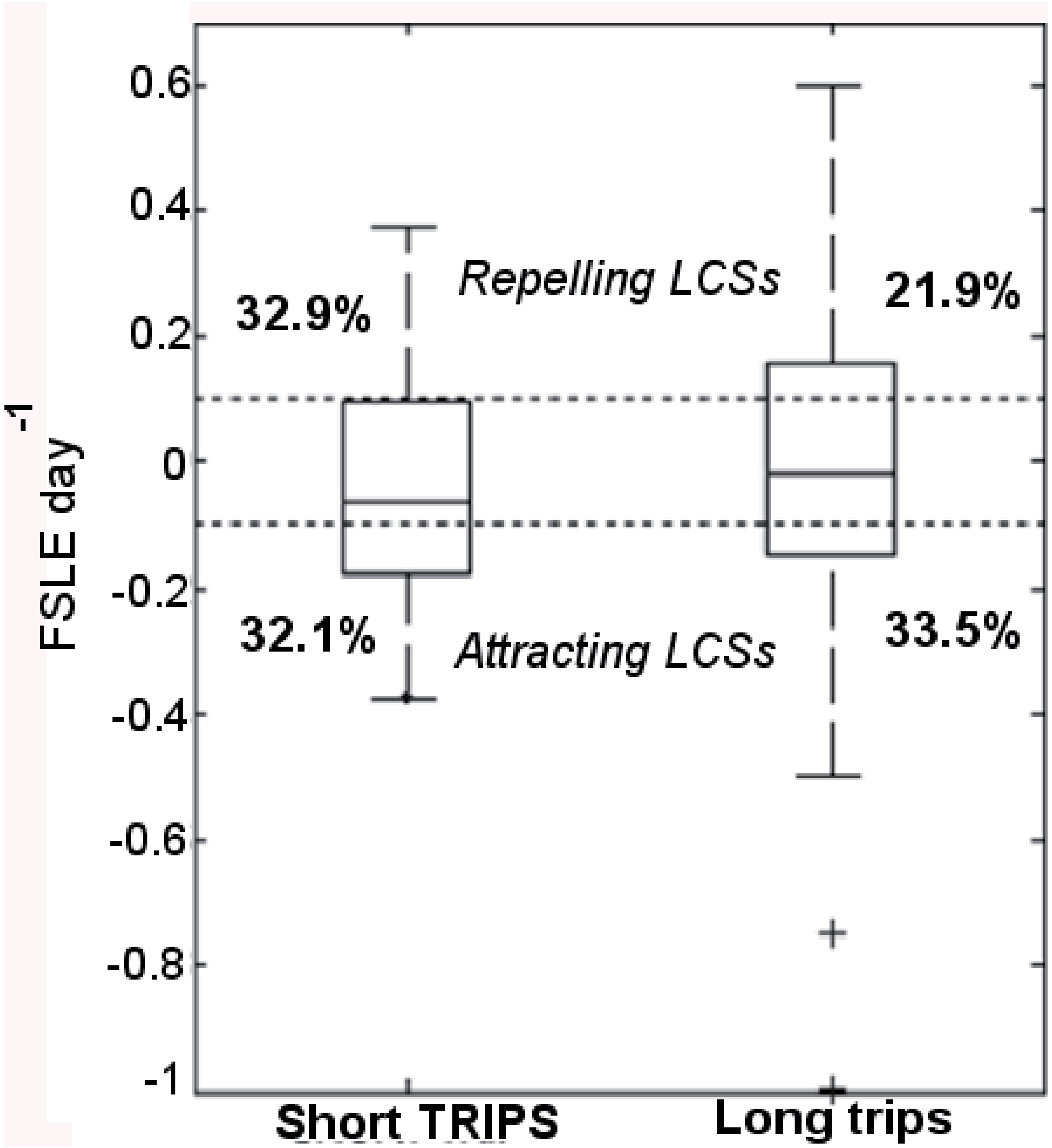

FIGURE 4

# Supporting Information

**SI Figure legend**

**Figure S1**

Figure S1: Box plots of the distribution of FSLEs during flying and foraging part of short and long trips. The upper and lower ends of the center box indicate the 75th and 25th percentiles of the data; the center of the box indicates the median. Suspected outliers appear in a box plot as individual points + outside the box. Dotted lines represent the threshold for detection of LCSs.

**Figure S2**

Figure S2: Box plots of the distribution of FSLEs. The upper and lower ends of the center box indicate the 75th and 25th percentiles of the data; the center of the box indicates the median. Suspected outliers appear in a box plot as individual points + outside the box. A) Outward and return part of short trips. B) Day and night short trips, C) outward and return part of long trips, D) day and night long trips. Dotted lines represent the threshold for detection of LCSs.

**Figure S3**

Figure S3: Comparison between the zonal and meridional components (cm/s) of the velocity field used in our study with those of Lagrangian buoy data in the same oceanographic region (Mozambique Channel). EGM currents are the sum of surface geostrophic anomalies (G), a climatological mean (M) and the surface Ekman velocity field (E). N is the number of data used for the comparison and the square of the correlation coefficient, $r^2$. In blue: all data points from Lagrangian drifters for our area of interest; in red: all points from Lagrangian

drifters for our area of interest when |Udrifter -Uegm| < 30cm/s and |Vdrifter - Vegm| < 30 cm/s.

## SI Table legend

**Table S1**

Table S1: Number of birds' positions at week t which are on the LCS of later weeks (t+i, i=1,3,5). The G-test statistics show a decreasing association between birds and LCSs as time lag between them increases.

**Table S2**

Table S2: Result of G-test statistics. Comparison between frequency of birds' positions on repelling or attracting LCS during flying and foraging and short and long trips; Alpha 5%.

**Table S3**

Table S3: Result of G-test statistics. Comparison between frequency of birds' positions on repelling or attracting LCS during outward and return part/day and night during short and long trips. Alpha 5%.

## SI text

*Methods*

### Great Frigatebirds

Seabirds' positions were interpolated to the same resolution of FSLEs. Because of Argos positioning errors and inherent errors in interpolating satellite data on a much finer grid, we say that a bird position is on a LCS if it is within a radius of 0.025° from a point where |FSLE|>0.1 $d^{-1}$. Following Weimerskirch et al. (1), trips were separated

in two categories, long and short ones. Typically Great Frigatebirds were doing long trips, mainly during incubation (58.5% of birds), when birds forage long distances from the colony, and shorter trips, mainly when they rear chicks (64.1%) and have to bring food regularly to the nest. A threshold at 617 km was used to distinguish both types of trips. 17 long trips and 33 short trips are separated and visualized on Figure 1. Short trips are located around the breeding colony in Europa Island and positions of long trips are mostly located in the western central part of the channel between 18°S and 26°S, except for 2 trips. Foraging patches were defined as the areas where flight speed between at least 3 successive Argos locations is lower than 10 km $h^{-1}$ (2). Therefore, only pairs of locations at sea separated by more than 30 min were used to limit erroneous estimates of speed because of the relative inaccuracy of the locations (1).

**Surface currents data**

The weekly global ¼° resolution product of surface currents developed by Sudre and Morrow (3) has been used over the time period January 1st, 2001 to December 31st, 2006. The surface currents are calculated from a combination of wind-driven Ekman currents, at 15 m depth, derived from Quikscat wind estimates, and geostrophic currents computed from time variable Sea Surface Heights. These SSH were calculated from mapped altimetric sea level anomalies combined with a mean dynamic topography from Rio et al (4). The weekly velocity data, which are then interpolated linearly to obtain a daily resolution with a 0.025° intergrid spacing, depend on the quality of their sources as the SSH fields and the scatterometer precision. However, they were validated with different types of *in situ* data such as Lagrangian drifting buoys, ADCP and current meter mooring data. In the Mozambique Channel (10°-30°S, 30°-50°E), zonal and meridional components of the velocity field show an average correlation with for e.g. Lagrangian buoy data between 0.71 and 0.76 (see Figure 3 SI).

When calculating the FSLEs from velocity data with a resolution of ¼ degree and interpolating down to 1/40° we are assuming that the small scale details of the velocity field are not important for the dispersion dynamics. This situation is called non-local dynamics (5) since it implies that the small scale transport is driven by the large scales. The assumption is correct for flows with an energy spectrum steepest than $k^{-3}$ which corresponds to 2D turbulence. Although there is some uncertainty in energy spectra for the marine surface, the calculations of Stammer (6) show that there is a decay of the energy spectra, at mid-latitudes, close to $k^{-3}$. Thus we might expect a weak sensitivity of FSLE computations of the surface ocean to the spatial resolution of the velocity field.

**Computation and analysis areas**

The full geographical area of the Mozambique Channel is used to make our numerical computations of FSLEs. We then defined our analysis areas large enough to cover the maximum extension of birds' trajectories and made the approximation to the closest proper rectangle fitting the best. Note that the computation areas are larger than the analysis ones, considering the fact that particles may leave the area before reaching the fixed prescribed final distance $\delta_f$.

**Statistical test Table S1:**

To compare the number of birds' positions at week t (from 1 to 10) which are on LCS at that given week, with the number of these birds' positions which fall on the LCS of different weeks t+i (i=1,2,…,9) we performed G-tests which quantify their independence. To do so, we consider all the seabirds' positions for a given week *t*. Then we compute the FSLE at week t, and identify which of the birds' positions correspond to LCS. Maintaining the original frigate positions at t, we compute the values of FSLEs for the whole time series of Lyapunov maps from t'=t to t'=t+i (i=1,2,…,9), identifying again which of the bird's positions are on LCSs. G-test were performed on these distributions of number of coincidences of LCSs at all times with the locations of birds at the given time t. Results are displayed on Table S1 for i=1,3,5 and show a decreasing association between birds and LCSs as time lag i between them increases.

Table1

| Positions at week t | on LCSs of week t | on LCSs of week t+1 | on LCSs of week t+3 | on LCSs of week t+5 |
|---|---|---|---|---|
| **WEEK1** | 19 | 14 | 9 | 21 |
| **WEEK2** | 55 | 49 | 34 | 56 |
| **WEEK4** | 146 | 106 | 106 | 99 |
| **WEEK5** | 137 | 114 | 112 | 118 |
| **WEEK6** | 89 | 69 | 89 | 81 |
| **WEEK7** | 72 | 67 | 81 | 71 |
| **WEEK8** | 53 | 50 | 41 | 28 |
| **WEEK9** | 61 | 59 | 48 | 66 |
| **WEEK10** | 45 | 28 | 46 | 48 |
| ***Gtest*** | | ***0.81*** | ***0.19*** | ***0.12*** |

Table2

| | | Flying | Foraging |
|---|---|---|---|
| Long trips | Repelling LCS FSLE>0.1 $day^{-1}$ | 318 | 50 |
| | Attracting LCS FSLE<-0.1 $day^{-1}$ | 333 | 37 |
| *G-test* | *N* | *738* | |
| | *G* | *2.29* | |
| | *p* | *0.13021* | |
| Short trips | Repelling LCS FSLE>0.1 $day^{-1}$ | 76 | 9 |
| | Attracting LCS FSLE<-0.1 $day^{-1}$ | 112 | 10 |
| *G-test* | *N* | 207 | |
| | *G* | *0.34* | |
| | *p* | *0.55993* | |

*Two tailed tests. Ho: seabirds share out equally on repelling and attracting structures when they fly or forage*

Table3

| | | OUTWARD | RETURN | DAY | NIGHT |
|---|---|---|---|---|---|
| Long trips | Repelling LCS FSLE>0.1 day$^{-1}$ | 196 | 156 | 188 | 162 |
| | Attracting LCS FSLE<-0.1 day$^{-1}$ | 186 | 165 | 164 | 181 |
| | *N* | *703* | | *695* | |
| | *G* | *0.513* | | *2.655* | |
| | *p* | *0.47395* | | *0.10325* | |
| short trips | Repelling LCS FSLE>0.1 day$^{-1}$ | 33 | 29 | 27 | 33 |
| | Attracting LCS FSLE<-0.1 day$^{-1}$ | 53 | 37 | 65 | 38 |
| | *N* | *152* | | *163* | |
| | *G* | *0.474* | | *5.003* | |
| | *p* | *0.49* | | ***0.0253*** | |

*Ho:seabirds share out equally on repelling and attracting structures during day and night and seabirds share out equally on repelling and attracting structures during outward and return flights*

Fig1

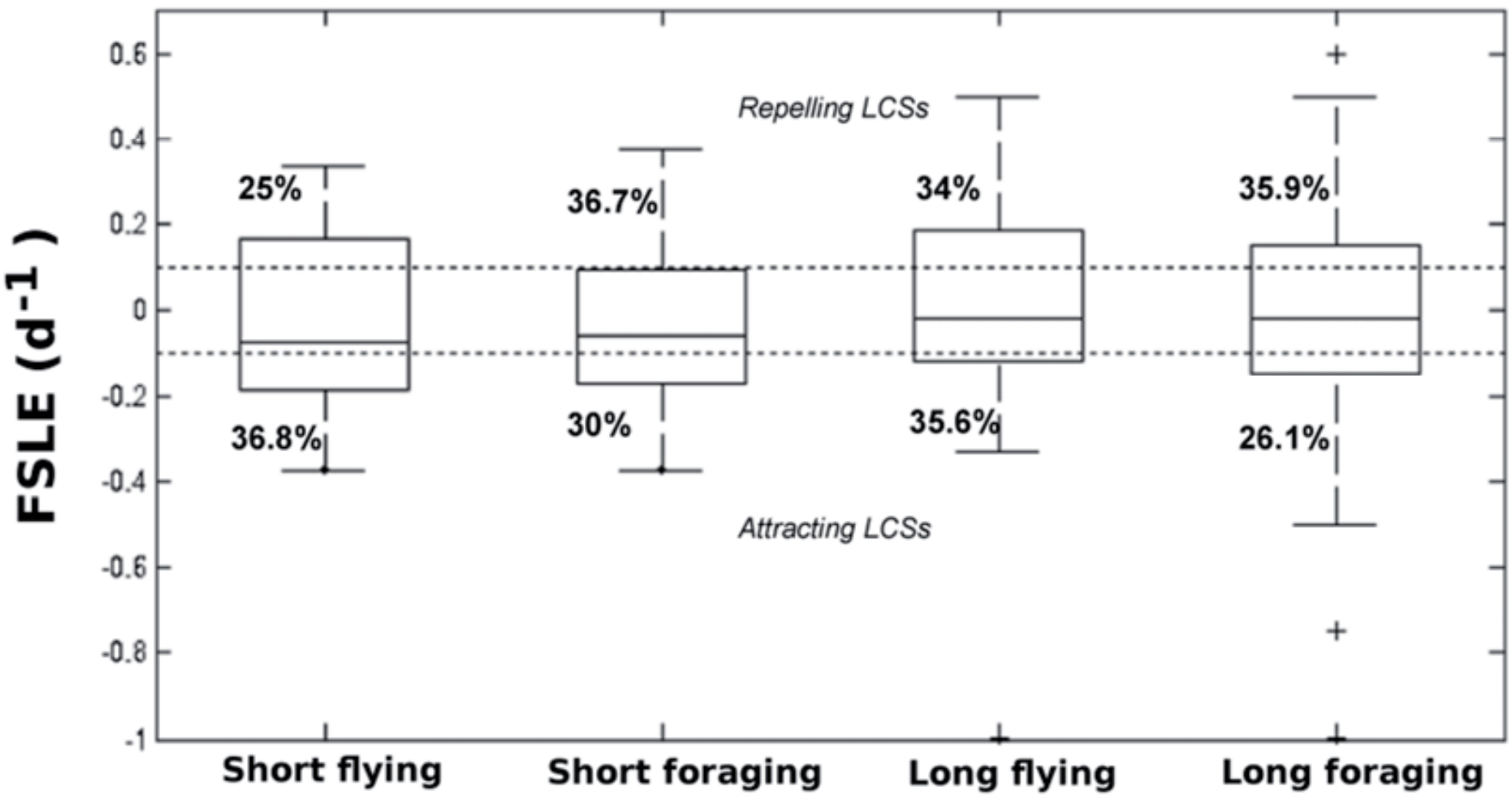

Fig2

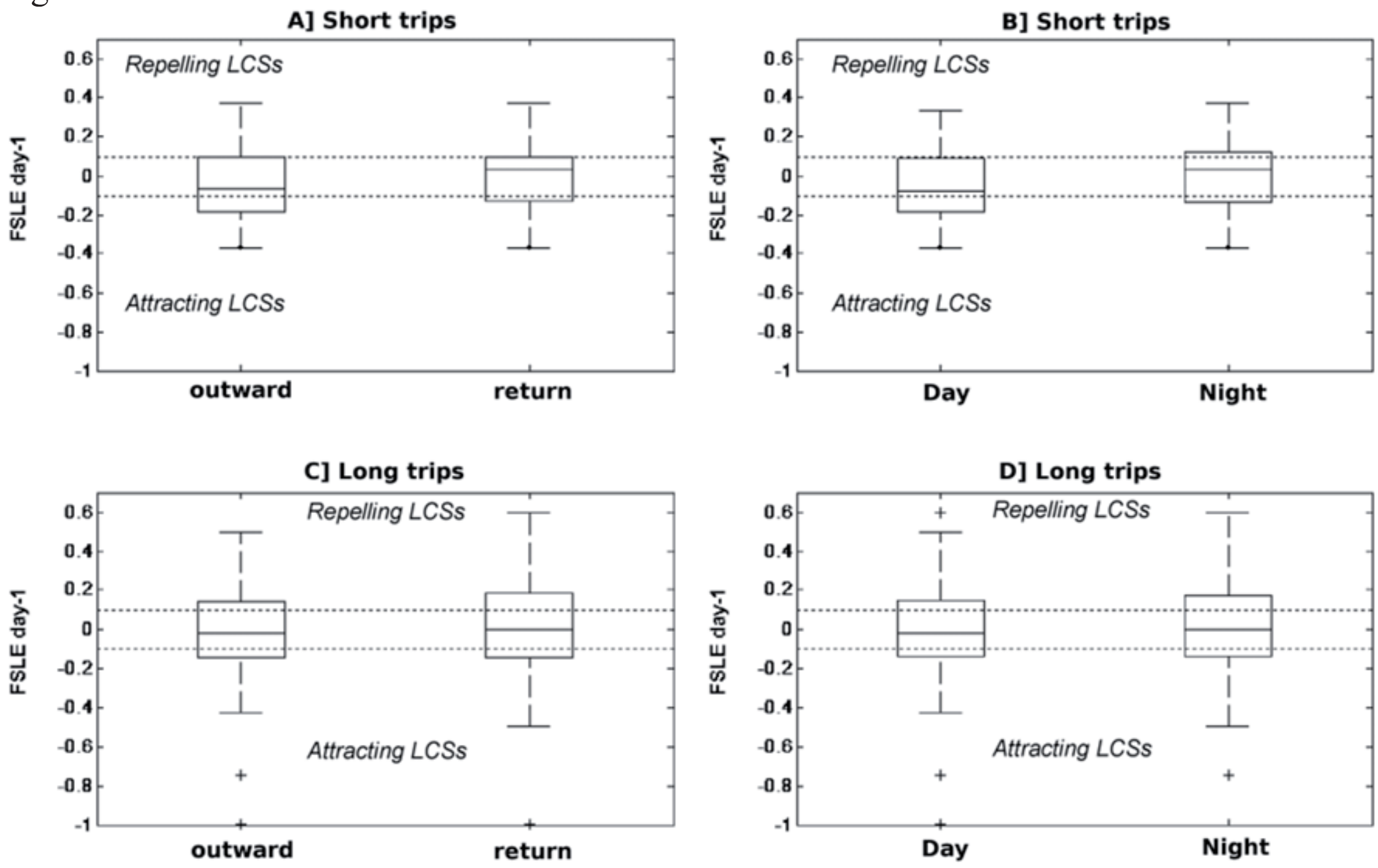

Fig3

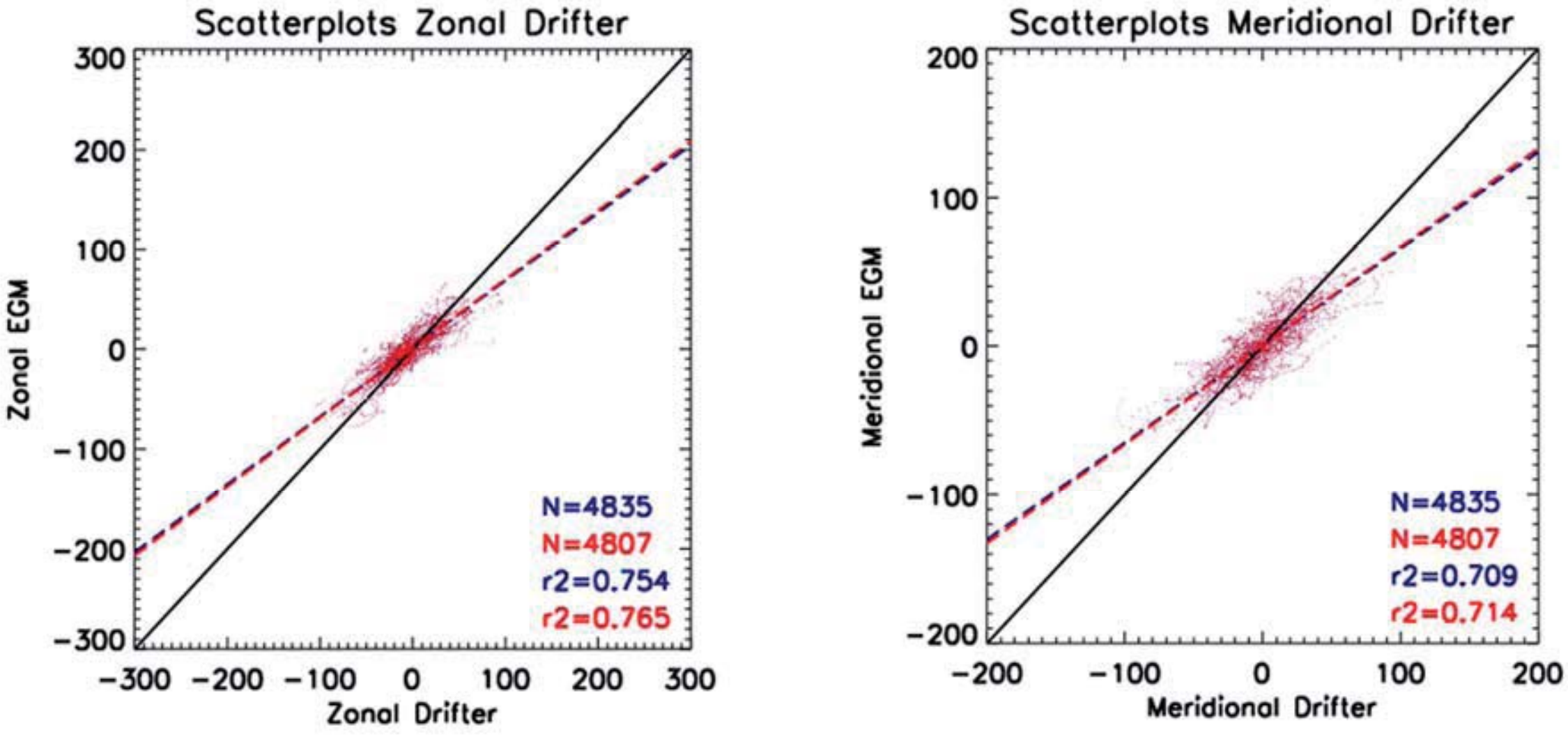